\documentstyle[12pt,preprint,aps]{revtex}
 
\begin{document}
\input{epsf}
 
\preprint{UMHEP-439}
\draft
\title{On dipenguin contribution to $D^0-\bar D^0$ mixing.}
\author{
Alexey A. Petrov}
\address{
Department of Physics and Astronomy \\
University of Massachusetts \\
Amherst MA 01003}
\maketitle
\begin{abstract}
\noindent
We study the impact of the four-quark dipenguin operator
to the $D^0-\bar D^0$ mixing.
It is shown to contribute to the short distance 
piece at the same order of 
magnitude as the box diagram. 
\end{abstract}

The phenomenon of meson mixing has been studied for a long time.
Observed in the $K^0-\bar {K^0}$ and $B^0-\bar {B^0}$
systems, it provides an 
extremely sensitive test of the Standard Model (SM) and essential for
the CP violation in the neutral meson system. 

It is well known that because of the 
GIM cancellation mechanism      
and large mass of the top quark the short distance
box diagram dominates in
$B^0-\bar {B^0}$ system and constitutes a significant
fraction of the $K^0-\bar {K^0}$ mixing amplitude.  
The case of $D^0-\bar {D^0}$ system is somewhat special: 
the $b$-quark contribution
to the fermion loop of the box diagram providing $\Delta C = 2$ 
transition is
diminished by a tiny $V_{ub}$ $CKM$ matrix element.
Thus, only the light quark mass difference guarantees
that mixing does take place.
The effect vanishes in the limit of the SU(3) invariance.
All of that results in the estimated 
value for $\Delta m_D$ being of the order of
$10^{-17} GeV$ if only short distance contributions are
taken into account \cite{datta}, \cite{disper}.

Calculating a box diagram and constructing the
effective Hamiltonian one realizes that 
the smallness of the short
distance piece is guaranteed by a factor of 
$(m_s^2-m_d^2)^2/M_W^2 m_c^2$ \cite{disper}:

\begin{eqnarray}
{\cal H}_{eff} = \frac{G_F}{\sqrt 2} 
\frac{\alpha}{8 \pi \sin ^2 \theta_W } \xi_s \xi_d 
\frac{(m_s^2-m_d^2)^2}{M_W^2 m_c^2} ~~~~~~~\nonumber \\
\Bigl ( \bar u \gamma_\mu (1 + \gamma_5) c
 \bar u \gamma_\mu (1 + \gamma_5) c +
2~  \bar u (1 - \gamma_5) c \bar 
u  (1 - \gamma_5) c \Bigr)
\end{eqnarray}
with $\xi_i = V^*_{ic} V_{iu}$. Here, the $b$-quark contribution is 
dropped. This leads to the following expression for the 
$\Delta m_D^{box}$ \cite{datta},\cite{disper}
\begin{equation} \label{msd}
\Delta m_D^{box}  =\frac{G_F}{\sqrt 2} 
\frac{\alpha}{4 \pi \sin ^2 \theta_W } \xi_s \xi_d 
\frac{4}{3} 
\frac{(m_s^2-m_d^2)^2}{M_W^2 m_c^2} f_D^2 m_D 
(B_D - 2 B_D')
\approx 0.5 \cdot 10^{-17} 
\Bigl[\frac{m_s}{0.2} \Bigr]^4
\Bigl[\frac{f_D}{f_\pi} \Bigr]^2
\end{equation}
with $f_\pi \simeq 132$ MeV, $f_D \simeq 165$ MeV, and 
$B_D=B_D'=1$ in the usual vacuum saturation approximation to
\begin{eqnarray}
\langle D^0 | O_1 | \bar D^0 \rangle =
\frac{8}{3} \frac{f_D^2 m_D^2}{2 m_D} B_D,~~~~~~ 
\langle D^0 | O_2 | \bar D^0 \rangle =
-\frac{5}{3} \frac{m_D^2}{m_c^2} 
\frac{f_D^2 m_D^2}{2 m_D} B_D' \nonumber \\
O_1=\bar u \gamma_\mu (1 + \gamma_5) c
 \bar u \gamma_\mu (1 + \gamma_5) c, ~~~~~
O_2=\bar u (1 - \gamma_5) c \bar 
u  (1 - \gamma_5) c ~~~
\end{eqnarray}
In contrast to the K-meson mixing, the appearance of the 
second operator can be traced to the fact that
mass of the external $c$-quark provides a large momentum 
scale. As one can see from the
Eq. (3.7) of \cite{datta}, the inclusion of the $b$-quark 
further decreases the box diagram contribution.

In this note we would like to address an additional contribution to
the short-distance D-meson mixing amplitude which is topologically
distinct from the box diagram - the so-called 
double penguin or ``dipenguin'' operator. This operator was 
initially introduced in \cite{donoghue} for $K^0-\bar {K^0}$
mixing amplitude
and has been subsequently studied in \cite{eeg} in
application to $K^0-\bar {K^0}$ as well as to  $B^0-\bar {B^0}$
systems. It was shown to be marginally important in the
former and completely negligible in the latter case.
Here we will introduce the dipenguin operator for the
$\Delta C = 2$ transitions. It will be shown that this operator 
contributes to the $D$-meson mass difference at the same order 
of magnitude as the usual box diagram.

The effective operator relevant to dipenguin $\Delta C = 2$ transition
can be obtained from the usual $\Delta C=1$ penguin vertex
(we neglect a tiny dipole contribution):

\begin{equation} \label{vertex}
\Gamma_\mu^a=
-\frac{G_F}{\sqrt 2} 
\frac{g_s}{4 \pi^2} F_1  
\bar u \gamma_\mu (1 + \gamma_5) \frac{\lambda^a}{2} c
(g^{\mu \nu} \partial^2 - \partial^\mu \partial^\nu )
A^a_\nu
\end{equation}
Here $F_1$ is a modified Inami-Lim function \cite{gerard}. 
Using unitarity of the CKM matrix it reads 
$F_1=\sum_i \xi_i F^i_1=\xi_s (F^s_1-F^d_1) +
\xi_b (F^b_1-F^d_1)$. It is common to 
discard $b$-quark contribution to $F_1$ as being suppressed
by small $V_{ub}$ factors. Note, however, that by keeping 
it we {\em enhance} the $F_1$ by approximately $20-30 \%$. Also,
for the intermediate $b$-quark this vertex (as well as the
following Hamiltonian) is {\em local}. From (\ref{vertex}) we 
obtain the following effective Hamiltonian
\begin{eqnarray} \label{dp}
{\cal H}_{dp}= 
-\frac{G_F^2}{128 \pi^2} 
\frac{\alpha_s}{\pi} F_1^2  
\Bigl (
(\bar u \gamma_\mu (1 + \gamma_5) \lambda^a c) 
\partial^\mu \partial^\nu
(\bar u \gamma_\nu (1 + \gamma_5) \lambda^a c) -
\nonumber \\
(\bar u \gamma_\mu (1 + \gamma_5) \lambda^a c)
\Box
(\bar u \gamma_\mu (1 + \gamma_5) \lambda^a c)
\Bigr )
\end{eqnarray}
In what follows, we denote two operators entering
(\ref{dp}) as $\tilde O_1$ and $\tilde O_2$.
In order to study the size of the dipenguin effects in charm mixing we
derive an estimate of the $\Delta m_D$ and compare with the usual box
diagram contribution. To do that, in addition to the usual vacuum
saturation approximation, we use pQCD in order to independently estimate 
the $\Delta m_D^{dp}$ by simply calculating the transition amplitude 
from Feynman diagram that determines the effect (Fig.1a). 
We believe that pQCD can provide a reliable order-of-magnitude
estimate of the dipenguin contribution since 
the momentum transferred through the gluon line in Fig. 1a is relatively
large, $Q^2 \sim -m_c^2$.  

Using the equations of motion and neglecting the up quark mass 
we obtain for the first operator
in (\ref{dp})

\begin{equation}
\tilde O_1 = \bar u \gamma_\mu (1 + \gamma_5) \lambda^a c 
~\partial^\mu \partial^\nu~
\bar u\gamma_\nu (1 + \gamma_5) \lambda^a c \simeq
m_c^2 ~\bar u (1 - \gamma_5) \lambda^a c ~
\bar u (1 - \gamma_5) \lambda^a c
\end{equation}
Before we compare the estimate of the $\Delta m_D^{dp}$ with the
relevant box diagram contribution we would like to note that the
dipenguin diagram does not have a power dependence upon the
internal quark masses. The leading behavior of the Inami-Lim function
is logarithmic in $m_{s(d)}$, and the estimate of the 
operator brings about power dependence upon the {\em external}
quark masses, i.e. $m_c$. This feature distiguishes this contribution 
from that of the usual box diagram.
Recalling the fact that the dominant contribution to 
$K$ and $B$ mixing amplitudes is proportional to the
square of the top quark mass, it is not surprising that
this effect is negligible in the K and B sectors.
It is the fact of the ``reduced'' heavy quark dependence of the
amplitude of $D^0-\bar {D^0}$ mixing which makes the
dipenguin operator contribution effectively enhanced.

Employing vacuum saturation method to estimate 
matrix elements we obtain

\begin{equation}
\langle D^0 | \tilde O_1 | \bar D^0 \rangle =
\frac{16}{9} \frac{f_D^2 m_D^4}{2 m_D} B , ~~~~~
\langle D^0 | \tilde O_2 | \bar D^0 \rangle =   
-\frac{32}{9} \frac{f_D^2 m_D^2 (2 m_c^2 - m_D^2)}{2 m_D} B'.
\end{equation}
with $B$ and $B'$ being the bag 
parameters. 
In addition to the vacuum saturation we assumed that
each derivative acting on the quark field involves
an average momentum of the quark.
This yields 

\begin{equation} \label{mdp}
\Delta m_D^{dp}=2 
\langle D^0 | {\cal H}^{dp} | \bar D^0 \rangle =
\frac{G_F^2}{72 \pi^2} \frac{\alpha_s}{\pi} 
F_1^2 (m_b^2,m_s^2,m_d^2)
f_D^2 m_D (m_D^2 - 4 m_c^2)
\end{equation}
This formula deserves some additional discussion.
In the case of $D^0-\bar D^0$ mixing the chief effect
comes from the light quark sector. This is true for
both the box and dipenguin diagrams, and makes the
calculation a little more involved - one
cannot simply discard the external quark momenta
(masses). That is why one must use a modified 
expression for the Inami-Lim function \cite{gerard}.
The leading contribution to $F_1^i$ comes from the
integral
\begin{equation} \label{il}
F_1^i(m_i^2, Q^2) = -4 \int_0^1
dx~ x(1-x) \ln \Bigl[ \frac{m_i^2}{M_W^2} -
\frac{Q^2}{M_W^2} x (1-x) 
\Bigr] = -4 \Bigl[ \frac{1}{6} \ln \frac{m_i^2}{M_W^2} +
\Pi \bigl(\frac{Q^2}{m_i^2} \bigr) \Bigr]
\end{equation}
where $ \Pi \bigl(\frac{Q^2}{m_i^2} \bigr)$ was defined in
\cite{gerard} for $Q^2 > 0$. In our case, of course, 
$Q^2 < 0 \simeq - m_c^2$.
The necessity of the second term becomes obvious if
one looks at the chiral limit, wherein the first logarithm in
(\ref{il}) blows up. The final result for
$F_1$ involves a delicate cancellation among these
contributions, yielding
a final result $|F_1| = 0.01-0.02$
(whereas $F_1 \sim O(1)$ if the momentum flow is discarded).
Comparing (\ref{msd}) and (\ref{mdp}) we find
\begin{equation}
|\frac{\Delta m_D^{dp}}{\Delta m_D^{box}}| \approx
\frac{\alpha_s}{8 \pi} \frac{F_1^2 (m_b^2,m_s^2,m_d^2) 
m_D^4}{| \xi_s \xi_d| (m_s^2-m_d^2)^2} 
\end{equation}
where we have put $m_c \approx m_D$, and $\alpha_s \simeq 0.4$.
The relative size of the box and dipenguin contribution 
shows that the latter is of the same order of magnitude 
as the box diagram (Our estimate
gives $\sim 20-50 \%$ depending on the choice of  
quark masses). This is not surprising 
if one recalls that higher order QCD corrections tend to 
``smooth out'' a power-like GIM suppression, just as
in the case of $B$-meson decays. This is not 
so relevant in the bottom (or strange) meson sector since
a large mass of the top quark actually converts
GIM-suppression to ``GIM-enhancement'', thus making higher
order corrections relatively unimportant. 

Another interesting observation is the fact that
the dipenguin diagram actually contributes to
the $\Delta m_D$ with a sign {\em opposite} to the
box diagram (compare (\ref{msd}) and (\ref{mdp}))!
This implies that the short-distance piece is even
smaller than was claimed in previous estimates based
solely on the box diagram contribution.
Note that this operator gives rise to a whole family of the 
diagrams (e.g. Fig. 1b) that by no means can 
be calculated in perturbative QCD but
might be potentially important for $D^0-\bar D^0$ mixing.

It must be stressed that $D^0-\bar {D^0}$ mixing is not
dominated by the short distance box diagram contribution 
\cite{datta} but rather
by  long distance pieces \cite{disper},
\cite{wolf}. This effect has been estimated in \cite{disper}
using dispersive techniques for a class of two-body pseudoscalar
intermediate states and was shown to boost the
value of $\Delta m_D$ to $\Delta m_D \sim 10^{-16}$ GeV.
Indeed, it is not excluded that additional contributions 
could conspire in a way that they cancel the two 
body piece \cite{georgi}.

In conclusion, we have estimated the contribution of the
dipenguin diagram to the short-distance amplitude for
$D^0-\bar D^0$ mixing. It is shown to   
contribute at the same order of magnitude as the box diagram.

{\bf Acknowledgments:} I would like to thank John F. Donoghue,
Eugene Golowich, and Barry Holstein for useful conversations 
and for reading the manuscript.

\section*{Appendix}

Here we will provide an independent pQCD estimate of the 
dipenguin matrix element $\langle D^0 | O^{dp} | 
\bar D^0 \rangle $. To do that we employ a Brodsky-Lepage
exclusive QCD description \cite{bl} for calculating the Feynman diagram
in Fig. 1a. In this formalism, the amplitude of
interest can be expressed as a convolution of the mesons'
distribution function with the hard scattering amplitude
$T(x,y)$. We assume that the momentum fraction carried by the 
c-quark in $D$ meson is $x$, and the momentum fraction 
carried by the $\bar c$ quark in the $\bar D$ meson is $1-y$.
As is well known, the heavy quark carries nearly all the 
momentum of the heavy-light bound state which makes 
the distribution functions
strongly peaked at $x \sim 1$ and $y \sim 0$.
This fact simplifies the choice of the form for the distribution 
amplitudes.
In order to obtain the estimate of the effect we use the simplest 
form for the distribution amplitudes
\begin{equation}
\phi(x) = \frac{f_D}{2 \sqrt{3}} \delta (1-x-\epsilon), ~~~
\phi(y) = \frac{f_D}{2 \sqrt{3}} \delta (\epsilon-y)
\end{equation}
These amplitudes are normalized such that
\begin{equation}
\int_0^1 \phi (x_1) dx_1 = \frac{f_D}{2 \sqrt{3}}, ~~~~
\langle 0 | A_\mu | \bar D^0 \rangle = i f_D {p_D}_\mu
\end{equation}
with $x_1 = 1-x$. The desired mass difference is then 
given by
\begin{equation} \label{ampl}
\Delta m_D^{dp} = 2 \langle D^0 |
{\cal H}_{dp} | \bar D^0 \rangle =
-\frac{1}{m_D}~ \int dx dy
\phi^*_D(y) T(x,y) \phi_D(y)
\end{equation}
where a ``scattering amplitude'' $T(x,y)$ can be read off the 
diagram shown in Fig. 1a and $m_D$ in the denominator comes from
the standard normalization of the meson states.
A computation of Eq. (\ref{ampl}) involves the effective vertex
$V_{8 \mu}$:
\begin{equation}
V_{8 \mu} = \frac{G_F}{\sqrt{2}} \frac{\lambda^a}{2}
\frac{g_s}{8 \pi^2}
\Bigl(F_1(Q^2) \Bigl[ Q^2 \gamma_\mu - Q_\mu \rlap/{Q} \Bigr]
(1 + \gamma_5) -
m_c F_2(Q^2) i \sigma_{\mu \nu} Q^\nu (1-\gamma_5)
\Bigr)
\end{equation}
where $F_i(Q^2) = \sum_{k} \xi_k F_{ki}(x_k,Q^2)$; $k=b,s,d$; 
$x_k=m^2_k/M^2_W$; $F_{ki}(x_k,Q^2)$ are the corresponding
modified Inami-Lim functions, and $Q^2$ is a momentum transferred 
through the gluon. Numerically $F_1=-0.015$ for
$m_b \simeq 5$ GeV, $m_s \simeq 0.2$ GeV, $m_d \simeq 0.01$ GeV.

The calculation of (\ref{ampl}) is relatively straightforward
and yields
\begin{equation} \label{mbl}
| \Delta m_D^{dp} | =\frac{G_F^2}{2 \pi^2} m_D^3 F_D^2 
\Bigl \{ \frac{C_F}{16} \frac{\alpha_s}{\pi} 
(F_1 - F_2)^2 \Bigr \}
\end{equation}
Here $C_F=4/3,~ \alpha_s \simeq 0.4$, 
and we put $m_c \approx m_D = 1.87$ GeV.  
The calculation amounts to $\Delta m_D^{dp} \simeq 
-0.2 \cdot 10^{-17} (f_D/f_\pi)^2 GeV$.
Note that the pQCD estimate and the vacuum saturation estimate
give reasonably close results.

\begin{figure}
\centerline{
\epsfbox{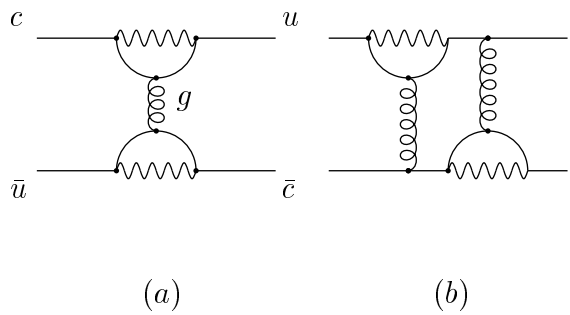}}
\caption{Dipenguin diagram (a) and a possible long-distance
contribution (b)}
\end{figure}

\end{document}